\documentclass[aps,prx,twocolumn,superscriptaddress,showpacs]{revtex4-1}
\usepackage{natbib}
\usepackage{graphicx}
\usepackage{amsmath}
\usepackage{bm}
\usepackage{color,soul}
\usepackage{hyperref}
\usepackage{float}
\usepackage{csquotes}
\usepackage{color,soul}
\hypersetup{colorlinks=true, urlcolor=blue, citecolor=blue}
\usepackage{braket}

\begin{document}
\title{ 
Effective tight-binding Hamiltonian for the
low-energy electronic structure of  the Cu-doped lead apatite and the parent compound}

\author{Mayank Gupta}
\affiliation{Condensed Matter Theory and Computational Lab, Department of Physics, IIT Madras, Chennai 600036, India}
\affiliation{Center for Atomistic Modelling and Materials Design, IIT Madras, Chennai 600036, India}
\author{S. Satpathy}
\email{satpathys@missouri.edu}
\affiliation{Condensed Matter Theory and Computational Lab, Department of Physics, IIT Madras, Chennai 600036, India}
\affiliation{Center for Atomistic Modelling and Materials Design, IIT Madras, Chennai 600036, India}
\affiliation{Department of Physics \& Astronomy, University of Missouri, Columbia, MO 65211, USA}  
\author{B. R. K. Nanda}
\email{nandab@iitm.ac.in}
\affiliation{Condensed Matter Theory and Computational Lab, Department of Physics, IIT Madras, Chennai 600036, India}
\affiliation{Center for Atomistic Modelling and Materials Design, IIT Madras, Chennai 600036, India}
 
\begin{abstract}

We examine the origin of the formation of narrow bands in LK-99 (Pb$_{9}$Cu(PO$_4$)$_6$O) and the parent compound without the Cu doping  using density functional theory calculations and model Hamiltonian studies. 
Explicit analytical expressions are given for a nearest-neighbor  tight-binding (TB) Hamiltonian in the momentum space for both the parent and the LK-99 compound, which can serve as an effective model to study various quantum phenomena including superconductivity. 
The parent material is an insulator with the buckle oxygen atom on the stacked triangular lattice forming the topmost bands, well-separated from the remaining oxygen band manifold. 
  The $C_3$ symmetry-driven two-band TB model  describes these two bands quite well. 
  These bands survive in the Cu-doped, LK-99, though with drastically altered band dispersion due to the Cu-O interaction. 
A similar two-band model involving the Cu $xz$ and $yz$ orbitals broadly describes the top two valence bands of LK-99. However, the band dispersions of both the Cu and O bands  are much better described 
by the  four-band TB model incorporating the Cu-O interactions on the buckled honeycomb lattice. 
We comment on the possible mechanisms of superconductivity in LK-99. even though the actual T$_c$ may be much smaller than reported, and suggest that interstitial Cu clusters leading to broad bands might have a role to play.


\end{abstract}  
\date{\today}					
\maketitle

\section {Introduction} 
A recent experimental study by Lee et al. \cite{lee2023roomtemperature,Lee} suggests that the lead apatite mineral LK-99 (Pb$_9$Cu(PO$_4$)$_6$O) exhibits room superconductivity under ambient pressure. As achieving superconductivity at room temperature has been a long desire since the discovery made by Kamerling Onnes in 1911, the present experimental revelation has created a lot of curiosity and intense theoretical and experimental investigations. Though the sharp drop in the resistivity and magnetic levitation have been demonstrated in the first experiment, the superconductivity in LK-99 is still debated due to the lack of sufficient data and the absence of zero resistivity. However, the possible existence of narrow bands around the Fermi level in this compound has formed an avenue to explore several exotic quantum phases. These include spin-liquid phase \cite{baskaran,sun2023metallization}, low-dimensional spin-exchange coupling \cite{Balents2010}, correlated electron behavior \cite{yue2023correlated,jiang2023pb9cupo46oh2}, possible Mott/charge-transfer insulating states \cite{baskaran}, topological quantum phases \cite{hirschmann2023tightbinding} emerging out of the $C_3$ and broken inversion symmetry of the crystal, and flat-band magnetism \cite{hirschmann2023tightbinding,jiang2023pb9cupo46oh2,held1}, etc. 

The excitement over realizing the room temperature superconductivity has led to many DFT calculations on LK-99 \cite{yue2023correlated, griffin,held1,kurleto2023pbapatite,jiang2023pb9cupo46oh2}. The salient feature of the electronic structure, which emerges from these studies, is that it has a pair of three-quarter-filled narrow bands dominating the Fermi level. Just below this pair, there is another pair of bands of similar dispersion. The orbital projected band structure studies reveal that, while all four bands have sizable O-$p$ characters, the upper two bands only have significant contributions from Cu-$d$ ($xz$ and $yz$) orbitals. Therefore, on the theoretical research front, the current attention is to developing model Hamiltonian that reproduces the four bands \cite{jiang2023pb9cupo46oh2,lee2023effective} and two band models \cite{hanbit,hirschmann2023tightbinding,tavakol2023minimal}, hence the electronic structure at the Fermi surface. Such models are helpful in examining the probable ground state quantum phases and proposing mechanisms that govern these phases.

\begin{figure*}
\includegraphics[scale=0.165]{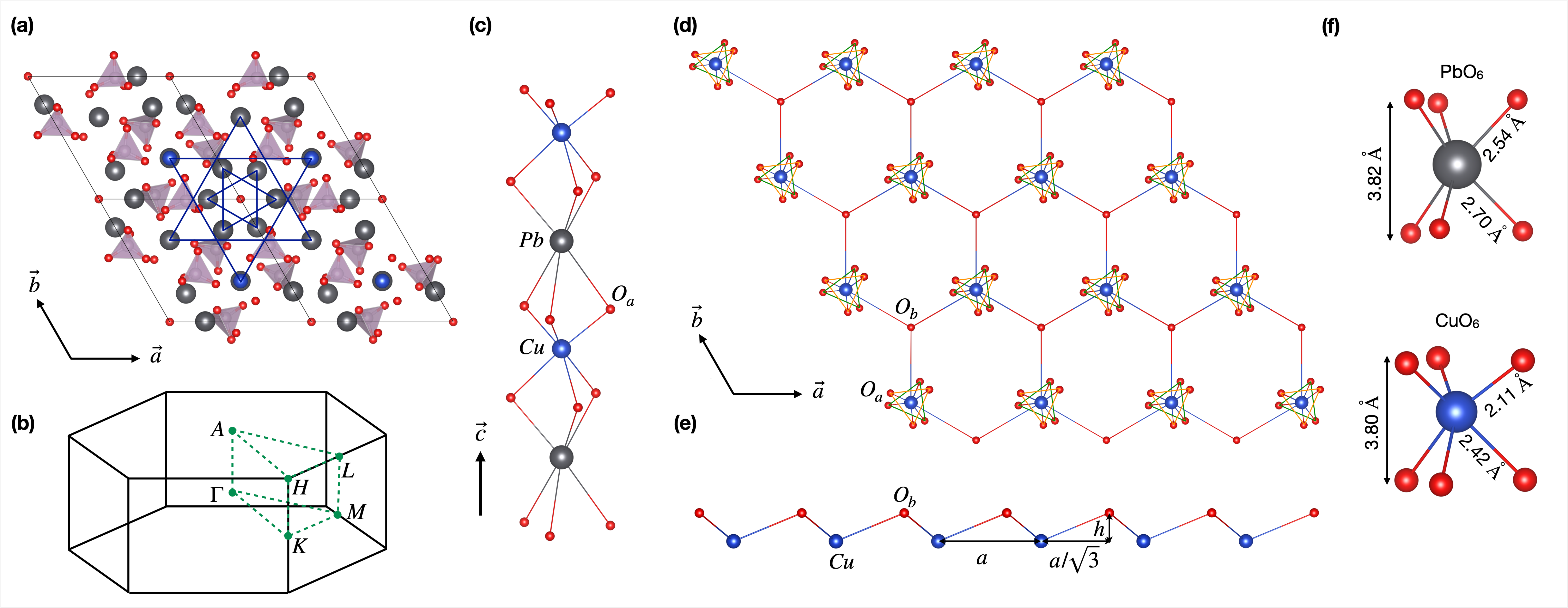}
\caption{Crystal structure of LK-99 (Pb$_9$Cu(PO$_4$)$_6$O) stabilizing in P3 space group.
(a) As viewed from the (001) plane. The concentric hexagonal rings of Pb and PO$_4$ tetrahedra pairs are highlighted. (b) The Brillouin zone. (c) Formation of a O-Pb-O-Cu-O-Pb-O chain along the z-axis. Each Cu (Pb) forms a CuO$_6$ (PbO$_6$) complex with a triangle of oxygen atoms up and a triangle of oxygen atoms down. We have named these oxygen atoms as apical oxygen (O$_a$).  
(d)  When seen from the (001) plane, the CuO$_6$ complex forms a triangular lattice, and with the lone oxygen, it forms a buckled Honeycomb lattice. We have named this oxygen as (O$_b$). It is buckled as O$_b$ has a vertical shift of $h$ ($\approx$ 2.3 \AA)  from the Cu plane as shown in (e). (f) The metal-oxygen bond lengths for PbO$_6$ and CuO$_6$ complexes imply that the latter is shrunk in volume.}
\label{structure}
\end{figure*}

A careful look at the proposed tight-binding (TB) models and the DFT studies suggests that the parent compound Pb$_{10}$(PO$_4$)$_6$O has been largely ignored in analyzing the electronic structure. Understanding the electronic structure of the parent compound is crucial to examining the host-Cu interaction. Furthermore, as far as TB models are concerned, few of them are two-band models with a minimal number of parameters that do not necessarily replicate the actual band structure. Some of these models have proposed symmetry-driven flat bands which do not appear in the actual band structure. Wannier formalism \cite{mao2023wannier} and NMTO downfolding methods \cite{yue2023correlated} are adopted to develop the four-band TB models. These models are governed by the in-plane $C_3$ symmetry for the Cu-O honeycomb lattice (see Fig. \ref{structure} for details). These models (a) do not provide a detailed insight into the role of CuO$_6$ complexes formed along the z-axis, informing the low-energy electronic structure, and (b) do not provide analytical parametric expressions in the k-space.


In this work, we have carried out DFT calculations and developed a set of TB model Hamiltonians (two-bands and four-bands) both for the parent compound and LK-99 to provide insight into (a) the formations of narrow bands and (b) the role of oxygen orbitals of CuO$_6$ complex as well as that of the Cu-O honeycomb lattice in governing the Cu-host interactions.  From the TB models, we have provided explicit analytical expressions of the Hamiltonian matrix elements in the $k-$space, which could be useful in studying the physics of parent and Cu-doped apatite. When optimized, the hopping parameters in these elements reproduce the actual band structure of the parent compound and LK-99 very well. Furthermore, these tunable parameters provide flexibility to develop mechanisms and phenomena. 

\section{Crystal structure of LK-99 and Computational details}
It has been proposed that the LK-99 inherits the crystal structure of the parent compound Pb$_{10}$(PO$_4$)$_6$O by replacing one of the Pb with Cu. The structure has P3 space group symmetry (No. 143). The structure is shown from different perspectives in Fig. \ref{structure}. The lattice parameters considered in this study are borrowed from the work of Lee et al. \cite{Lee} ($a$ = $b$ =  9.8 \AA \hspace{0.1cm} and $c$ = 7.6 \AA).

The salient features of the LK-99 structure are: (I)  When projected onto the (001) plane, the parent compound has an inner Pb hexagonal ring and concentrically an outer Pb-Cu hexagonal ring formed by Pb The paired PO$_4$ tetrahedral to form a hexagonal ring by occupying the interstitial space of the outer ring (see Fig. \ref{structure} (a)). (II) As shown in Fig. \ref{structure} (c), along the $c$-axis, an O-Pb-O-Cu-O-Pb-O chain is formed where each Cu(Pb) ion forms a non-octahedral Cu(Pb)O$_6$ complex. The complex has two triangles of apical oxygen atoms (O$_a$), one on the top of Cu(Pb) and one on the bottom. The upper and lower triangles are rotated with respect to each other by an angle of $\pi$/6. Each of the apical oxygen is associated with on of the PO$_4$ tetrahedra ( see Fig. \ref{structure} (a) ).  (III) As can be seen from Fig. \ref{structure} (d),  in the $ab$ plane, the CuO$_6$ complex forms a triangular lattice, and with the non-tetrahedral/buckled oxygen (O$_b$), it forms a buckled honeycomb lattice. It is buckled as O$_b$ has a vertical shift of around 2.3 \AA \hspace{0.1cm} from the Cu plane, as shown in  Fig. \ref{structure} (e). The triangular symmetry is deterministic in the electronic structure, as will be evident from our TB model discussed in Section V. (IV) The substitution of Cu brings significant local distortions. It is demonstrated in Fig. \ref{structure} (f) through the bond-length variation of the PbO$_6$ and CuO$_6$ complexes. As an average, the bond lengths are shrunk by 0.6 \AA. (V) the paired PO$_4$ tetrahedra form a Kagome lattice (not shown in the figure). However, the Kagome symmetry is less important as we will see that the low-energy electronic structure of LK-99 is hardly influenced by the electrons of these strongly ionic complexes. 

To describe the electronic structure of the parent compound and LK-99, we have carried out density-functional theory calculations and have developed a set of  two-band and four-band TB models. The DFT calculations are performed using the pseudopotential based projector-augmented wave (PAW) method \cite{Blochl1994,Kresse1999} within the framework of PBE-GGA exchange-correlation functional using the Vienna ab-initio Simulation Package (VASP) \cite{Kresse1996}. A plane wave cutoff of 450 eV and an 8$\times$8$\times$8 $\Gamma$-centered k-mesh is used for BZ integration. We find that the spin-orbit coupling does not influence the electronic structure significantly. Hence, it is ignored in our results and analysis. The Wannierization is done by using the Wannier-90 package \cite{MOSTOFI2008685}. 
The TB model is discussed in Sections V and VI.

\begin{figure}
\includegraphics[scale=0.2]{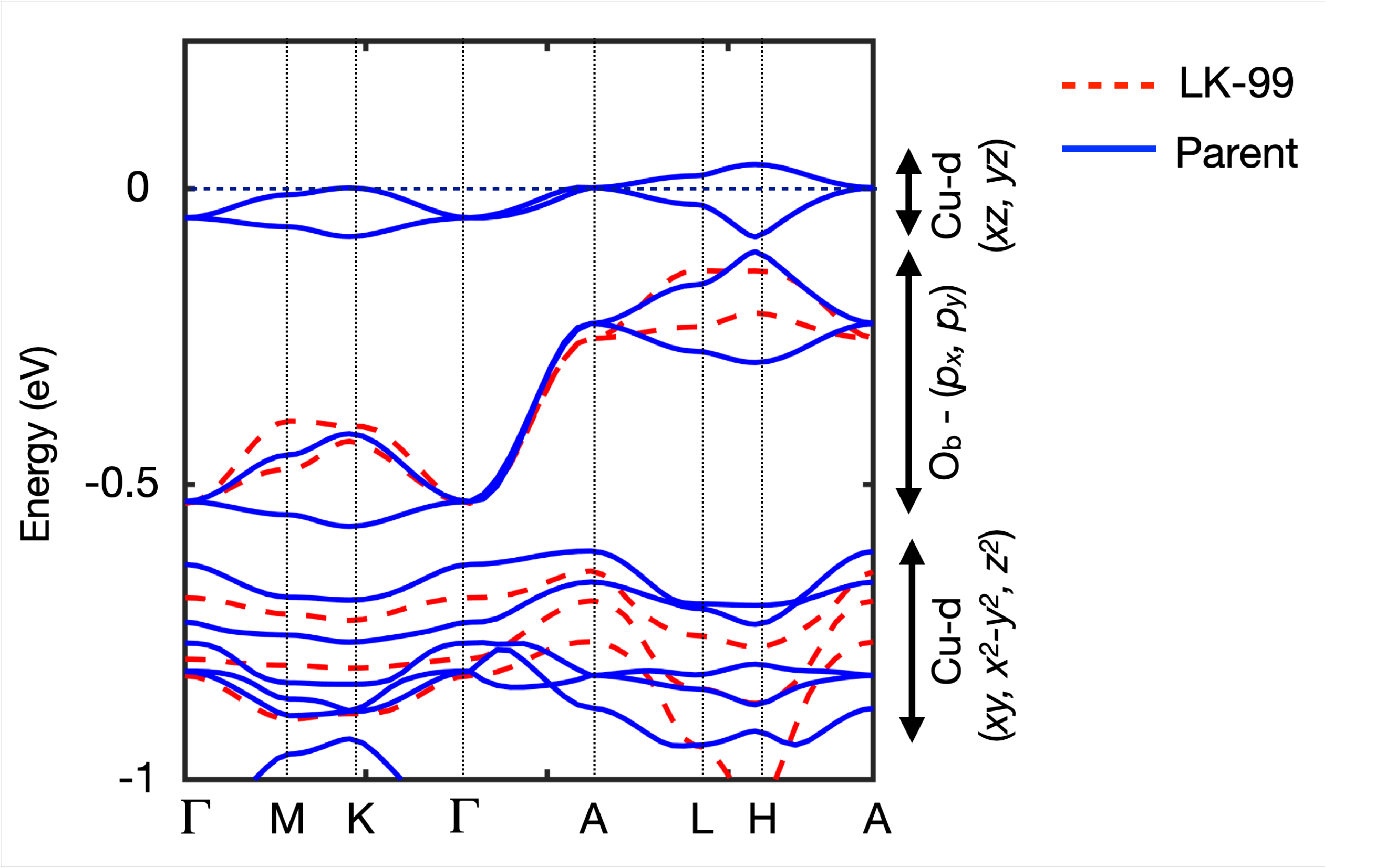}
\caption{The band structure of LK-99 (solid lines) and the parent compound Pb$_{10}$(PO$_4$)$_6$O (dashed lines). The dominance of the orbital characters in different energy spectra is indicated.}
\label{dft-band-parent}
\end{figure}

\section{Electronic Structure of the parent compound and LK-99}

The band structure of the parent compound Pb$_{10}$(PO$_4$)$_6$O is shown in Fig. \ref{dft-band-parent} (red dashed lines). In this compound, the trivalent phosphate polyanion induces 2+ charge state in Pb, which results in a 6$s^2$ closed-shell electronic configuration. Consequently, the bands dominated by Pb-$s$ characters lie around 6.5 eV below the Fermi level (not shown here), while the Pb-$p$ states occupy the conduction band spectrum. The O$_a$-$p$ and O$_b$-$p$ states occupy the valence band spectrum. It is important to note that the lattice formed by the O$_b$ atoms have $C_3$ symmetry. In this symmetry, the  $p$ orbitals segregate into the pair ($p_x$, $p_y$) and $p_z$. Accordingly, from the band structure, we observe that there is a pair of valence bands  lying close to the Fermi level and are well separated from the manifold of the rest of the valence bands formed by the O-$p_x$ and $p_y$ orbitals. These two orbitals can intermix such that under three-fold rotations, the state occurs a phase $e^{2n\pi/3}$ ($n$ is an integer). The bands with $p_z$ character is buried inside the valence band are invariant under rotation.

\begin{figure}
\includegraphics[scale=0.18]{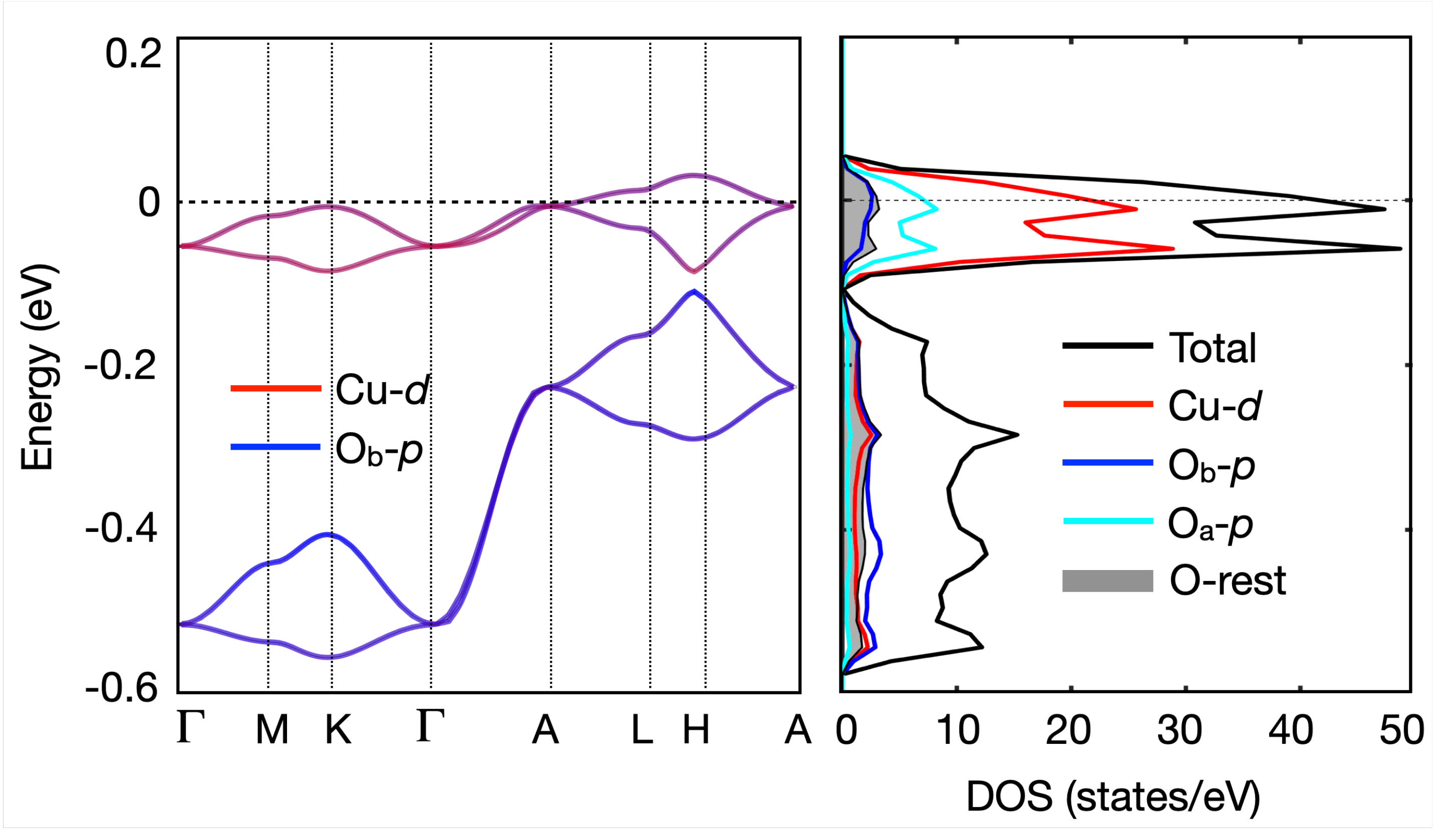}
\caption{(left) Visual representation of the orbital characters in the top four valence bands of LK-99. (right) Total and orbtial resolved partial DOS corresponding to bands on the left.}
\label{band-dos}
\end{figure}

With one of the Pb atoms is replaced by Cu, as shown in Fig. \ref{structure}, Cu forms a triangular lattice as O$_b$ does. Therefore, under the C$_3$ symmetry, the five-fold degenerate orbitals, which are quadratic in nature, are segregated to three different groups 3$z^2 - r^2$, which is invariant under three-fold rotation, ($x^2$ - $y^2$, $xy$), and ($xz$, $yz$). The members of the latter two pairs, like the case of O$_b$($p_x$, $p_y$),  can intermix such that they produce a phase $e^{2n\pi/3}$ under three-fold rotations. Furthermore, the trigonal crystal field formed by the apical oxygen atoms makes these three groups non-degenerate in energy. From our DFT calculations, while we find a narrow split in energy between ($xy$, $x^2-y^2$) and $z^2$ bands, the pair of bands formed by ($xz$, $yz$) orbitals are well separated and lie on the Fermi level. These two bands are three-quarters filled. This is due to the fact that Cu is in 2+ charge state leading to d$^9$$s^0$ configuration, and six of the $d$-electrons are now occupied by ($xy$, $x^2-y^2$) and $z^2$ states. The electronic structure of LK-99 presented here is in agreement with the recently reported DFT results \cite{yue2023correlated, griffin,held1,kurleto2023pbapatite,jiang2023pb9cupo46oh2}.

Further analysis of the band structure of the parent compound and LK-99 reveals the following two important aspects. First, in both cases, the bands are narrow across all the energy windows. The bandwidth of the bands lie in the range of 0.1 to 0.3 eV. Similar observations are made by Jena and Nanda in the case of transition metal based Olivine phosphates \cite{Jena2016}. The formation of narrow bands in the Olivine phosphates is attributed to the fact that the oxygen atoms have a strong affinity to form  the ionic PO$_4$ tetrahedral complexes. These complexes are formed at a cost of significant distortion to the metal-oxygen complexes, which in turn lower the symmetry of the crystal significantly \cite{PARIDA2018133}. Furthermore, these metal-oxygen complexes are well separated from each other, with the PO$_4$ complexes occupying the interstitial regions. As a consequence, the covalent hybridization among the valence states becomes very weak, which leads to the formation of narrow bands in these crystals. Such an analysis can be extended to the apatite minerals as well.

Second, when the band structure of the parent compound and LK-99 are compared, the striking difference that one observes is that the presence of Cu has altered the shape of the dispersion of the  O$_b$-($p_x$, $p_y$) bands significantly.  This implies that the Cu$-d$ states initiate  new covalent bonds with the O$_b-p$ states. To gain more insight into it, in Fig. \ref{band-dos}, we have plotted the atom and orbital resolved band structure and DOS. The visible mix of the colors confirms that the shapes of the upper two bands and lower two bands are influenced by the formation of covalent bonds Cu$-d$ and O$_b-p$. The DOS shown in Fig. \ref{band-dos} doubly verifies this inference as we find that the upper two bands, though are predominantly Cu$-d$, as expected, have a reasonable contribution from the O$_b-p$ orbitals. Interestingly, we observe that the O$_a-p$ orbitals too collectively contribute significantly to the upper two bands which suggests the mixing of the Cu$-d$ and O$_p$ states within the CuO$_6$ complex. The TB models developed in the next two sections will give insight into the nature and strength of the covalent interactions occurring in the parent compound and LK-99.

\begin{table}[!htb]
  \caption{Orbital characters of the top two bands in the parent compound
  Pb$_{10}$(PO$_4$)$_6$O.  O$_b$ denotes the lone buckle O atom, while O$_a$ denotes the apical oxygen (six in number). }
    \centering
    \begin{tabular}{c  c  c c c  c c}
    \hline
    O$_b$ (x/y)   &    O$_b$ (z)     & O$_a$ & O (rest) & Pb  & P & Total  \\
     \hline
0.34 &   0.01 & 0.11 & 0.22 & 0.32 & 0.01 & 1.00 \\
      \hline
    \end{tabular} 
    \label{parent-characters}
\end{table}

\section{Model Hamiltonian for the parent material}
The electronic structure analysis from the DFT calculations, made in the last section,  infers that the top two valence bands of the parent compound and the top two distinct pair of bands in LK-99 form the low-energy electronic structure and determine the probable existence of the quantum states (e.g. superconductivity, flat-band magnetism, spin-liquid, correlated Mott insulator, etc.) in them. Therefore, it is instructive to examine the formation of these bands and for this purpose, we have developed a set of TB models in this section and the next section.



\begin{figure}
    \centering
    \includegraphics[scale=0.2]{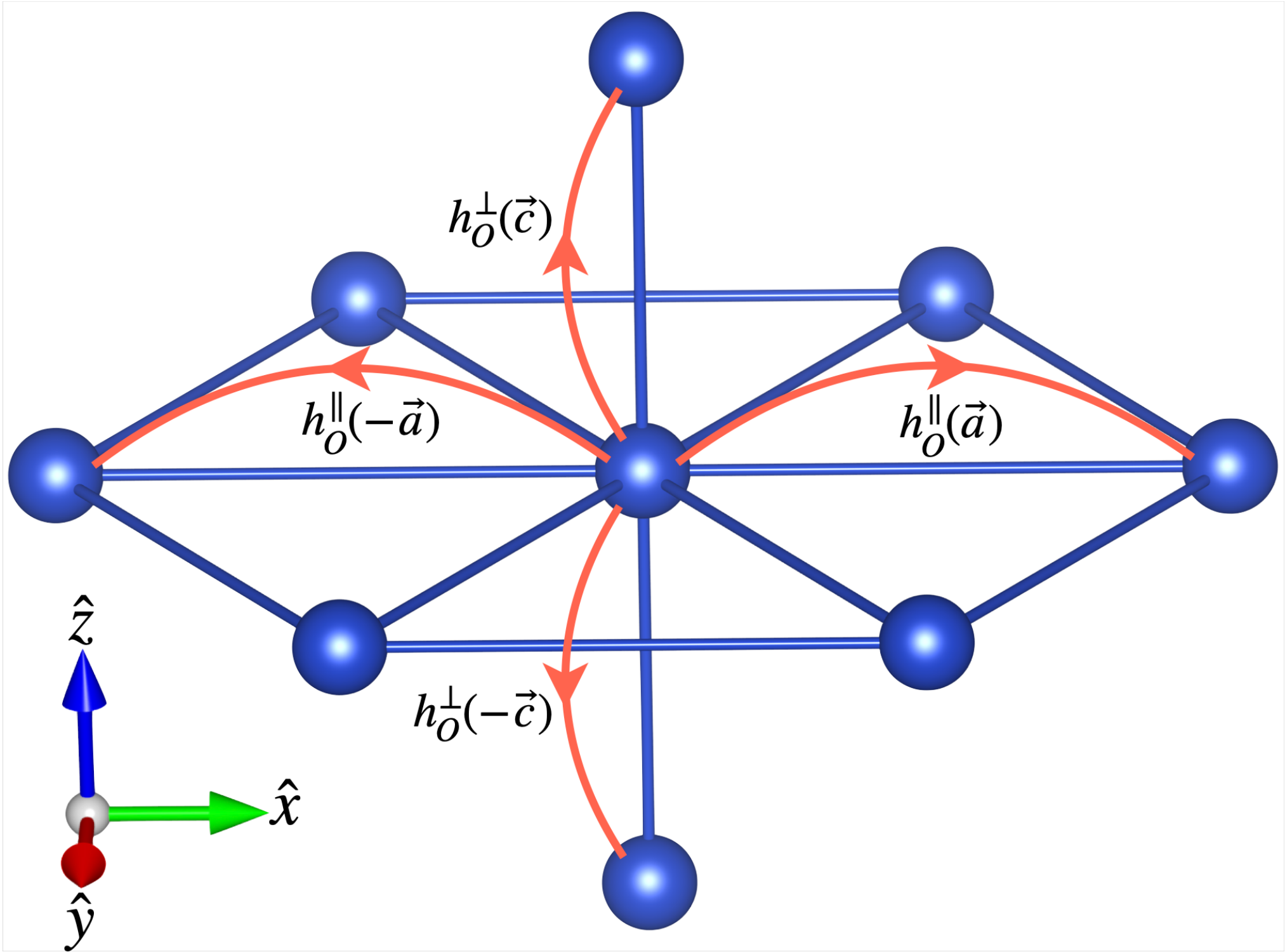}
    \caption{Triangular lattice of the O$_b$ 
    atoms and the hopping matrices
    in the parent compound. The remaining hopping matrices between other NNs are obtained by $C_3$ rotations.}
    \label{fig:parent-hopping}
\end{figure}

The orbital characters of the top two bands are shown in Table \ref{parent-characters},
from which it is clear that these bands are derived from the O$_b$-
($p_x$, $p_y$) orbitals. The loan O$_b$ atom contributes as much as 34 \%, 
the six apical oxygen $O_a$ contribute 11 \% together (or 2 \% each), the ten Pb atoms
contribute 3\% each, while the P atoms contribute a negligible amount.
This analysis therefore suggests the following
two-band TB model based on the $p_x / p_y$ orbitals of O$_b$ in the unit cell:

%
\begin{equation}
H = \sum_{i \alpha}\varepsilon_p  c^\dagger_{i\alpha}  c_{i\alpha} 
+ \sum_{\substack{
<ij> \\ \alpha \beta} } h_O^\perp c^\dagger_{i\alpha}  c_{j\beta}
+ \sum_{\substack{
<ij> \\ \alpha \beta} }  h_O^\parallel c^\dagger_{i\alpha}  c_{j\beta} + h. c.,
\end{equation}
where $c^\dagger_{i\alpha}$ creates an electron at site  $i$ and orbital $\alpha \ (\alpha = p_x, p_y)$, on the triangular O$_b$ lattice, and $<ij>$ indicates summation over distinct pairs of nearest neighbors (NN). The hopping elements are schematically shown in Fig. \ref{fig:parent-hopping}. In the $(p_x, p_y)$ subspace,   $h_O^\perp$ 
is a $2 \times 2$ matrix, representing hopping along the perpendicular direction while  
$h_O^\parallel$ is the hopping along the planar direction.       
The hopping matrices $h_O^\parallel$ along the three NNs in the $xy$-plane are not independent; rather, they are related to each other via the $C_3$ rotational symmetry of the structure. Thus by knowing the hopping matrix along $\vec{a}$ and $-\vec{a}$, we can compute the corresponding matrices along all six directions connecting NN O-O atoms, using the transformation relation
$
    h_O^\parallel (\theta_\gamma) = R^T(\theta_\gamma)h_O^\parallel (0)   R(\theta_\gamma),
$
where 
\begin{equation}
R (\theta) = 
\begin{pmatrix}
\cos \theta  \hspace{2mm}     -  \sin \theta \\
 \sin \theta  \hspace{2mm}    \cos \theta
    \end{pmatrix},
\end{equation}
is the rotation matrix for rotation about the $\hat z$ axis for both  ($x, y$) and  ($xz, yz$) sets of orbitals. The form of the hopping matrices along a NN can be obtained by considering the symmetry of the structure, or by explicit downfolding of a minimal set of orbitals that carries the symmetry of the structure, or by examining the TB matrix elements from the DFT methods such as the NMTO method or the Wannier functions. The matrices take the following general form:
\begin{eqnarray}
h_O^\parallel (\vec a) & = & 
\begin{pmatrix}
t_1 \hspace{2mm}       t_4  \\
t_3 \hspace{2mm}       t_2 \\
    \end{pmatrix},
     \hspace{5mm}
h_O^\parallel (-\vec a) = 
\begin{pmatrix}
t_1 \hspace{2mm}       t_3  \\
t_4 \hspace{2mm}       t_2 \\
    \end{pmatrix}, \nonumber  \\
h_O^\perp (\vec c) & = & 
\begin{pmatrix}
t_5 \hspace{2mm}       t_6  \\
-t_6 \hspace{2mm}       t_5 \\
    \end{pmatrix},
     \hspace{5mm}
h_O^\perp (-\vec c) = 
\begin{pmatrix}
t_5 \hspace{2mm}       -t_6  \\
t_6 \hspace{2mm}       t_5 \\
    \end{pmatrix}.
\end{eqnarray}

The Hamiltonian in the Bloch function basis
\begin {equation}
d^\dagger_{\Vec{k}m} = N^{-1/2}\sum_i e^{i\vec{k}\cdot \vec{R_i}}c^\dagger_{im},
\label{BF}
\end{equation}
where the sum is over the positions of all $m$ orbitals, is given by

\begin{eqnarray}
h(\vec k) & = & 
\begin{pmatrix}
h_{xx}  \hspace{2mm}      h_{xy}  \\
h_{xy}^*  \hspace{2mm}      h_{yy}  \\
    \end{pmatrix},
    \label{HO}
\end{eqnarray}
where the matrix elements are given by
%
%
%
%
%

\begin{figure}
    \centering
    \includegraphics[scale=0.18]{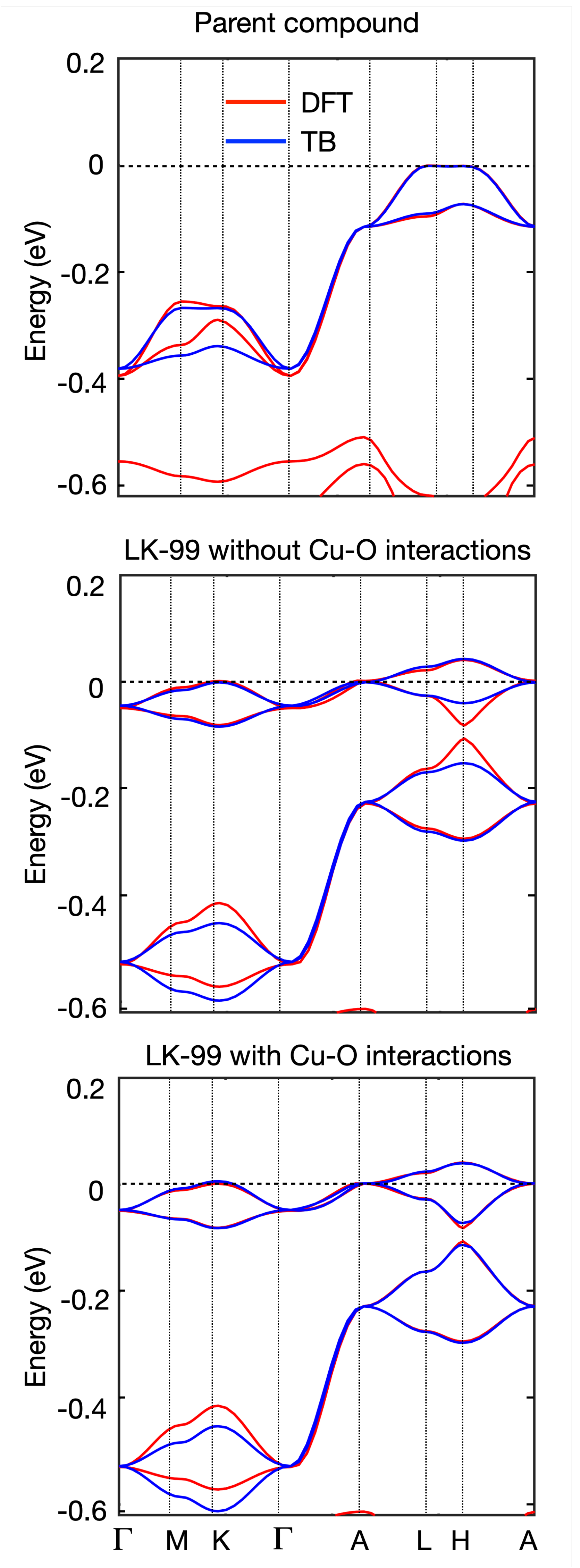}
    \caption{DFT vs. TB bands for the parent compound (a), for the Cu bands in LK-99 from the  two-band model (b), and  for the four Cu/O bands from the four-band model (c).}
    \label{TB-DFT-parent} 
\end{figure}

\begin{widetext}
\begin{eqnarray}
h_{xx} (\vec k) &=& 2 t_1 \cos(k_x a)+(t_1 + 3 t_2) c_x c_y 
-  \sqrt 3 (t_3 + t_4) s_x s_y + 2 t_5  \cos(k_z c), \nonumber \\
%
%
%
h_{yy}  (\vec k)  &=& 2 t_2 \cos(k_x a)+(3 t_1  + t_2) c_x c_y +  \sqrt 3  (t_3 + t_4 ) s_x s_y+2 t_5  \cos(k_z c),  \nonumber \\
h_{xy}  (\vec k)  &=&   (t_3 + t_4) (\cos (k_x a) - c_x c_y) 
 +   i (t_3 - t_4 ) (2 s_x c_y - \sin (k_x a))
 +  \sqrt 3  (t_1  - t_2) s_xs_y+2 i t_6  \sin(k_z c),  
 \end{eqnarray}
\end{widetext} 
where $c_x=\cos (k_x a /2)$, $c_y=\cos (\sqrt 3 k_y b /2)$, $s_x=\sin (k_x a /2)$, and $s_y=\sin (\sqrt 3 k_y b /2)$.
We have fitted the DFT band structure of the parent compound with this model, with the Hamiltonian parameters listed in Table \ref{parent-parameters} and the resulting TB band structure is compared with
the DFT bands in Fig. \ref{TB-DFT-parent}. The agreement is quite good, given the simplicity of the model. The agreement can be improved if desired by retaining further-neighbor interactions.
%

\section{Model Hamiltonian for Cu substituted Pb-apatite (LK-99)}
In this section, we construct a nearest-neighbor (NN) TB model Hamiltonian for 
the top four bands, the top two being ``Cu" bands, while the next two ``O" bands. 
Wannier functions computed from DFT indicate that the top two bands are
dominated by the Cu-$d_{xz}$ and $d_{yz}$ orbitals, while the next two bands below are dominated by the O$_b$-$p_x$ and $p_y$ orbitals, just like in the parent compound, where O$_b$ denotes the ``buckle" oxygen atom in the unit cell that along with the Cu atom forms the buckled honeycomb lattice (see Fig. \ref{structure}).  
 
 Further insights come from the  orbital characters of these bands, which are  summarized in Table \ref{lk99-characters}. As the Table shows, the top two bands have the majority of character (57\%) from the Cu atom, but also a significant character from the six apical oxygens O$_a$ (19 \%) 
 as well as from the buckle oxygen O$_b$ (7 \%). The next two bands below in energy again have a large 25\% character from the lone O$_b$, but a large component comes from the 
 Cu orbitals as well, indicating a strong hybridization between Cu and O$_b$. The rest 53\% character is from the remaining twenty-plus atoms, or $\sim 2$\% per atom, which is significant, but relatively small. 
 This justifies the description of these four bands in terms of the Cu-$d_{xz}$ and $d_{yz}$ orbitals together with the $p_x$ and $p_y$ orbitals of the buckle O$_b$ atom. 
 
\begin{table}[!htb]
  \caption{Hopping parameters (in meV) for the buckle oxygen atom O$_b$ for the parent compound.
}
    \centering
    \begin{tabular}{c  c  c c   c c c}
    \hline
    $t_1$ & $t_2$&$t_3$&$t_4$&$t_5$&$t_6$ &$\epsilon_O$\\
     \hline
-19.6 & 2.2 &  -9.1 & 4.6 & -66.5 & 1.1  & -197.0\\
      \hline
    \end{tabular} 
    \label{parent-parameters}
\end{table}

\begin{figure}
    \centering
    \includegraphics[scale=0.2]{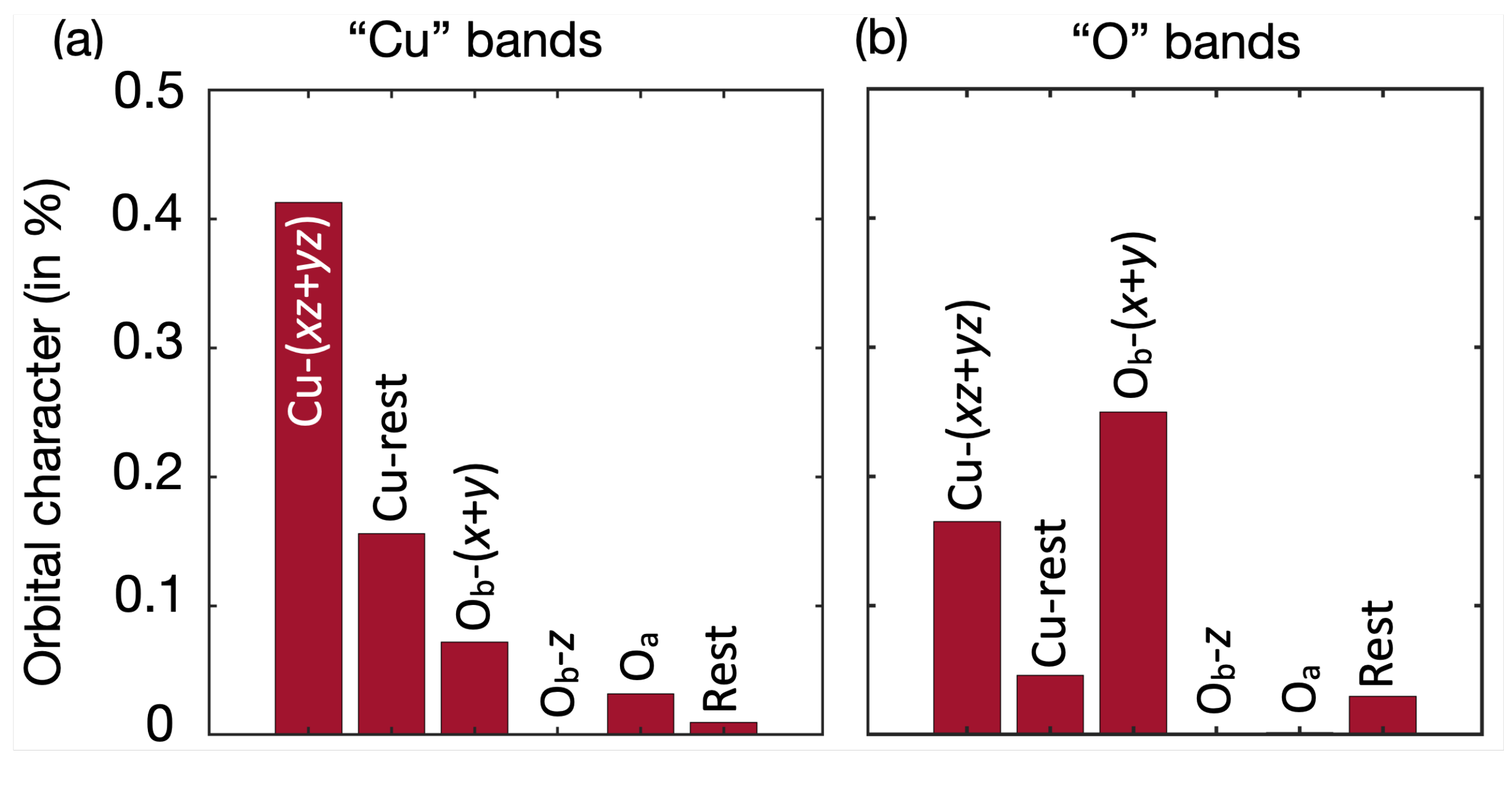}
    \caption{Bar chart representation of the contribution from different orbitals in forming (a) the top two valence bands (``Cu" bands) and next two bands (``O" bands). The bar chart is based on Table \ref{lk99-characters}.}
    \label{occu-bar}
\end{figure}

\begin{table}[!htb]
  \caption{Orbital characters of the topmost four bands in LK-99, derived from Cu and the buckle O$_b$
  atom.  }
    \centering
    \begin{tabular}{c | c  c c c c c c }
    \hline\hline
    Bands & Cu &Cu& O$_b$  & O$_b$ & O$_a$ & Rest & Total  \\
& $xz$/$yz$   & rest $d$ & $x/y$ & $z$   \\
     \hline
      ``Cu" bands&0.41 &0.16   & 0.07 & 0.00  & 0.19 & 0.17 & 1.00\\
      ``O" bands&0.17 & 0.05  & 0.25 &  0.00 &  0.01 & 0.53 & 1.00\\
      \hline\hline
    \end{tabular} 
    \label{lk99-characters}
\end{table}

The NN hopping matrix elements between the O$_b$ or the Cu orbitals have the same form 
as Eq. (3) in the parent compound, since the system continues to have the C$_3$ point group symmetry. For the oxygen hopping, we keep the same parameters $t_1, t_2, ..., t_6$,
while we use the primed quantities for the Cu-Cu hopping. We thus write the NN hopping matrices for Cu along the plane and normal to the plane as
\begin{eqnarray}
h_{Cu}^\parallel (\vec a) & = & 
\begin{pmatrix}
t'_1 \hspace{2mm}       t'_4  \\
t'_3 \hspace{2mm}       t'_2 \\
    \end{pmatrix},
     \hspace{5mm}
h_{Cu}^\parallel (-\vec a) = 
\begin{pmatrix}
t'_1 \hspace{2mm}       t'_3  \\
t'_4 \hspace{2mm}       t'_2 \\
    \end{pmatrix}, \nonumber  \\
h_{Cu}^\perp (\vec c) & = & 
\begin{pmatrix}
t'_5 \hspace{2mm}       t'_6  \\
-t'_6 \hspace{2mm}       t'_5 \\
    \end{pmatrix},
     \hspace{5mm}
h_{Cu}^\perp (-\vec c) = 
\begin{pmatrix}
t'_5 \hspace{2mm}       -t'_6  \\  
t'_6 \hspace{2mm}       t'_5 \\
    \end{pmatrix}.
\end{eqnarray}
The new terms are the Cu-O hopping, which are along the three NN Cu-O bonds on the plane,
connected by the $C_3$ symmetry. Actually, these Cu-O bonds are not exactly on the plane; rather, they form a buckled honeycomb structure as indicated in Fig. \ref{structure} (d). 
Along $\hat z$, we have an O$_b$ chain, so that there are no Cu-O NN along the perpendicular direction. With the Cu at the origin and O$_b$ at the position $\vec r = (0, 3^{-1/2} a, s)$, where $s \approx 2.31$ \AA\ is the height of the O$_b$ atoms above the Cu plane in the buckled honeycomb structure, the Cu-O hopping matrix is

\begin{equation}
T (\vec r) \equiv \bra {Cu, \vec 0} H \ket {O_b, \vec r}= 
\begin{pmatrix}
t_1'' \hspace{2mm}       t_3''  \\  
t_4'' \hspace{2mm}      t_2'' \\
    \end{pmatrix},
\end{equation}
with the basis set in the sequence: Cu-$d_{xz}$, Cu-$d_{yz}$, O$_b$-p$_x$,
and O$_b$-p$_y$.
In the Bloch function basis $(d^\dagger_{xz}, d^\dagger_{yz}, d^\dagger_{x}$, and $d^\dagger_{y})$, the $4 \times 4$ Hamiltonian becomes 
\begin{equation}
H (\vec k) = 
\begin{pmatrix}
H_{\rm Cu} \hspace{2mm}       H_{\rm Cu-O} \\   
H_{\rm Cu-O}^*  \hspace{2mm}      H_{\rm O} \\
    \end{pmatrix},
\end{equation}
where $H_{\rm O}$ is given by Eq. (\ref{HO}), $H_{\rm Cu}$ is exactly the same  except 
the parameters are different ($t_1', t_2', ..., t_6'$), and $H_{\rm Cu-O}$ is the Cu-O interaction
term 
\begin{equation}
H_{\rm Cu-O}  = 
\begin{pmatrix}
h_{xz, x} \hspace{2mm}       h_{xz, y} \\  
h_{yz, x}  \hspace{2mm}      h_{yz, y} \\
    \end{pmatrix},
\end{equation}
%
\begin{eqnarray}
   h_{xz, x} & = &  \frac{1}{2} \big[ 2t_1'' e^{i \phi_y}  +(t_1''+3 t_2'') c_x+ \sqrt{3}i (t_3''+t_4'') s'_x  \big], \nonumber \\
h_{xz, y} & = & \frac{1}{2} \big[ 2t_3''   e^{i \phi_y}  +(t_{3}''-3\,t_4'')c_x-\sqrt{3}i (t_1''-t_2'')  s_x' \big], \nonumber \\
h_{yz, x}  & = &   \frac{1}{2} \big[ 2t_{4}'' e^{i \phi_y} +(t_4''  -3 t_3'') c_x
   - \sqrt 3 i (t_1''-t_2'') s'_x  \big],  \nonumber \\
h_{yz, y} & = &  \frac{1}{2}  \big[ 2t_{2}'' e^{i \phi_y} +(t_2''+ 3  t_1'')c_x 
   - \sqrt 3 i  (t_3''+t_4'')s'_x  \big],
     \label{H-Cu-O}
\end{eqnarray}
%
%
where $c_x = \cos (k_x a /2)$
as defined before, $s'_x = \sin(k_xa/2\sqrt 3)$, and $\phi_y = k_yb/2\sqrt{3}$.
The parameters for LK99 obtained by fitting the DFT band structure with the $4 \times 4$ TB Hamiltonian are listed in Table \ref{lk99-parameters}.

Note that in Eq. (\ref{H-Cu-O}), a phase factor $e^{i\gamma} =  \exp[{i(k_zs-k_yb/2\sqrt 3)}]$ that multiplied all four expressions on the right-hand side of the equalities  has been omitted. This phase factor comes because we implicitly defined all TB orbitals in Eq. (\ref{BF}) with the ``exp($i\vec{k}\cdot$($\vec{R}$ + $\vec{\tau}$))" phase factor, where $\tau$ is the position of the corresponding atomic orbital in the unit cell. If the phase factor $e^{i\gamma}$ is kept in Eq. (\ref{H-Cu-O}), all orbitals in the basis contain the phase factor $e^{i \vec k \cdot \tau}$. Thus, taking the Cu atom as the origin, the Cu-$xz$ and $yz$ orbitals in the basis have no phase factors, while the O- $x$ and $y$ orbitals would have the phase factor $e^{i \vec k \cdot \tau_O}$
(which is precisely $e^{i\gamma}$ defined above),
where $\tau_O$ is the position of the O$_b$ atom with respect to the Cu atom. If the phase factor is omitted as we have done in Eq. (\ref{H-Cu-O}), all four atomic orbitals in the basis are defined without any phase factors. The energies obtained from the Hamiltonian are obviously the same in both cases.

The parameters for LK99 obtained by fitting the DFT band structure with the $4 \times 4$ TB Hamiltonian are listed in Table \ref{lk99-parameters}.

\begin{table}[!htb]
  \caption{Hopping parameters (in meV) for LK-99. Oxygen parameters are denoted by unprimed, Cu by single primed, and the Cu-O interactions by double-primed quantities.}
    \centering
    \begin{tabular}{c |c  c  c c   c c c}
        \hline

Cu &      $t'_1$ & $t'_2$&$t'_3$&$t'_4$&$t'_5$&$t'_6$ & $\varepsilon_d$\\   
           \hline
& -7.0 & 2.2 &  -14.1 & -1.7 & -2.2 & 2.5 &  0 \\
      \hline
       O & $t_1$ & $t_2$&$t_3$&$t_4$&$t_5$&$t_6$ & $\varepsilon_p$ \\
     \hline
& 2.1 & 1.0 &  1.6 & -25.2 & -85.2 & 2.4 & -29.3\\
      \hline
  Cu-O &           $t^{''}_1$ & $t^{''}_2$&$t^{''}_3$&$t^{''}_4$ \\
  \hline
& -4.0 & -46.0 &  57.2 & 9.5 \\
      \hline         
    \end{tabular} 
    \label{lk99-parameters}
\end{table}

\section {Discussions}

Even though the initial observation of the room-temperature superconductivity has failed to be replicated, it is very likely that LK-99 is still a superconductor, perhaps with a much lower T$_c$. However, with the half-filled narrow Cu bands with an $U/W \sim 30$ ($U$ $\approx$ 3 eV \cite{yue2023correlated}, W $\approx$ 0.1 eV), the system should be an anti-ferromagnetic Mott insulator, contrary to the observation of diamagnetism and superconductivity.  This suggests that the observed behavior could possibly come from off-stoichiometric samples, which would lead to a doped Mott insulator. 
There is a wide body of literature on superconductivity in doped Mott insulators in connection with the cuprate superconductors. The doped holes in the Cu bands could produce Zhang-Rice singlets, analogous to the cuprates \cite{Zhang-rice}. The RVB mechanism and also the t-J model developed for the cuprates suggest a $T_c \sim xJ$, where $x$ is the hole concentration and $J \sim t^2/U $. \cite{BASKARAN1987,Baskaran1987PRL,Baskaran1988,Kivelson1987}
Applied to LK-99, with $x = 10\%$, this would however produce a very small  T$_c \sim 20$ mK.

In view of this, the idea of a broad-band doped Mott insulator has been proposed \cite{baskaran}. Band calculations (including our results presented above) generally assume that the Cu goes into the lead substitutional site for the Pb chains running along the c direction, resulting in a very narrow Cu d bands at the Fermi energy. However, the exact position of the Cu dopants is as of now  unknown,
and it is conceivable that Cu goes into  interstitial sites in addition to the substitutional sites on the Pb chain,
forming a random network of weakly connected clusters and chains of Cu$^0$. In fact, it has already been pointed out that the LK-99 samples might be inhomogeneous, containing superconducting droplets embedded in a non-superconducting matrix \cite{abramian2023remarks}. Our test band calculations with a chain of Cu atoms replacing the O$_b$ chain along the c axis does reveal a broad Cu (s) band with bandwidth $W \approx 2-3$ eV [ M. Gupta, G. Baskaran, et al. (to be published)]. However, it requires three Cu atoms in lieu of one O$_b$ atom, which means the concentration of Cu would have to be very high. For lower Cu concentration, one could have pockets of such interstitial Cu chains or clusters of Cu, which could serve as an RVB reservoir, and a Josephson-type pair tunneling would lead to high-T$_c$.  Though T$_c$ would be high, it would be a weak superconductor with low critical current. Broadband RVB in quasi 2D and with a large effective exchange $J$  has been suggested for MgB$_2$ \cite{Baskaran2002,Black-Schaffe}. Also, the broadband low dimensional idea has been used  \cite{baskaran2019singlet,Baskaran2022}  to explain the reported granular and hot superconductivity in Ag-Au nanoclusters \cite{Saha_2022}. 

The idea of the flat-band superconductivity, of the type that has been proposed for twisted graphene, is not relevant here because the bands are narrow on account of the smallness of the TB hopping integrals, rather than destructive interference as the electron propagates \cite{Hwang2021}, which happens, e.g., in the much-studied Kagom\'e lattice. In the limit of perfectly flat bands due to localized atomic states, there would be no conductivity, let alone superconductivity. 

Thus it is difficult to reconcile the room-temperature superconductivity in the LK-99 and the observed diamagnetic behavior with the existing theories. Accurate determination of the structure for the Cu-doped system is urgently to make further progress on the understanding of the electronic structure and evaluation of potential mechanisms for room-temperature superconductivity. Imperfect, inhomogeneous crystals will not necessarily be detrimental to superconductivity but might even be the cause. 
In addition to superconductivity, the lead apatites are bound to have other interesting phenomena arising from the richness of their crystal structure.

\section {Summary and conclusion}

To summarize, we have comprehensively investigated the electronic structure of Cu-doped Pb apatite by carrying out density functional theory calculations and a series of model Hamiltonian studies on the Bloch function basis. The models are based on the Cu-Cu, Cu-O, and O-O nearest-neighbor interactions and the $C_3$ symmetry of the lattice in ab-plane. From our analysis, we find that the four-band valence manifold, which occupies the vicinity of the Fermi level, is three-dimensional in nature. This manifold is predominantly formed by the interactions within a buckled honeycomb lattice made out of an oxygen triangular sub-lattice and a CuO$_6$ triangular sub-lattice. From the model, we have developed explicit analytical expressions of the Hamiltonian matrix elements in the k-space. These expressions with tunable hopping parameters could be useful to develop new mechanisms and phenomena. We argue that the formation narrow band should be attributed to the presence of strongly bonded PO$_4$ tetrahedral ionic complexes. These complexes are formed by distorting the metal-oxygen complexes in the system and keeping them well separated. As a result, the covalent hybridization becomes weak to form narrow bands. The narrow bands are not the answer to the expected room temperature superconductivity as they can produce a T$_c$ of the order of 20 mK. However, if Cu is doped at the Pb site as has been assumed in the present study in the literature, these narrow bands become a rich source to exhibit correlated quantum phenomena. 

\textit{Acknowledgements:} We thank Prof. G. Baskaran for many insightful discussions. This work is funded by the Department of Science and Technology, India, through Grant No. CRG/2020/004330. SS thanks SERB India for the Fulbright fellowship. MG and BRKN acknowledge the support of HPCE, IIT Madras for providing computational facilities.

{\it Note added:} While preparing this manuscript, we came across several papers \cite{jiang2023pb9cupo46oh2,lee2023effective,hanbit,hirschmann2023tightbinding,tavakol2023minimal} that discussed the electronic structure of the LK-99 from DFT calculations, which broadly agree with our results. An important new point of our paper is the explicit analytical expression for the tight-binding Hamiltonian, which could be helpful for further modeling both the undoped and the Cu-doped lead apatite.

\end{document}